\documentstyle[prb,aps,amsfonts,floats,twocolumn,epsfig]{revtex}
  
\begin{document}
\title{How to mesh up Ewald sums (II): \\ An accurate error 
estimate for the P$^{3}$M algorithm}
\author{Markus Deserno, Christian Holm}
\address{Max-Planck-Institut f{\"u}r Polymerforschung, 
Ackermannweg 10, 55128 Mainz, Germany}
\date{\today}
\maketitle



\newcommand{\erfc}{{\mbox{erfc}}}
\newcommand{\infd}{{\mbox{d}}}
\newcommand{\mod}{{\mbox{{~}mod{~}}}}

\newcommand{\MM}{{\Bbb{M}}}
\newcommand{\NN}{{\Bbb{N}}}
\newcommand{\RR}{{\Bbb{R}}}
\newcommand{\ZZ}{{\Bbb{Z}}}

\newcommand{\VECd}{{\mathbf{d}}}
\newcommand{\VECe}{{\mathbf{e}}}
\newcommand{\VECk}{{\mathbf{k}}}
\newcommand{\VECl}{{\mathbf{l}}}
\newcommand{\VECm}{{\mathbf{m}}}
\newcommand{\VECn}{{\mathbf{n}}}
\newcommand{\VECr}{{\mathbf{r}}}
\newcommand{\VECD}{{\mathbf{D}}}
\newcommand{\VECE}{{\mathbf{E}}}
\newcommand{\VECF}{{\mathbf{F}}}
\newcommand{\VECR}{{\mathbf{R}}}
\newcommand{\VECchi}{{\mbox{\boldmath$\chi$}}}

\newcommand{\CALC}{{\mathcal C}}
\newcommand{\CALI}{{\mathcal I}}
\newcommand{\CALL}{{\mathcal L}}
\newcommand{\CALM}{{\mathcal M}}
\newcommand{\CALQ}{{\mathcal Q}}
\newcommand{\CALR}{{\mathcal R}}

\newcommand{\exa}{{^{\mbox{\footnotesize exa}}}}
\newcommand{\maxi}{{_{\mbox{\scriptsize max}}}}
\newcommand{\Mesh}{{_{\mbox{\scriptsize M}}}}
\newcommand{\nd}{{^{\mbox{\footnotesize nd}}}}
\newcommand{\opt}{{_{\mbox{\footnotesize opt}}}}
\newcommand{\self}{{^{\mbox{\scriptsize self}}}}
\newcommand{\th}{{^{\mbox{\footnotesize th}}}}

\newcommand{\D}{\displaystyle}


\hspace{2cm}\begin{abstract}
\begin{centering}
\parbox{16cm}{
\vspace{-1cm}
\hfill
\parbox{14.2cm}{
We construct an accurate estimate for the root mean square force error of 
the par\-ti\-cle-par\-ti\-cle--par\-ti\-cle-mesh (P$^{3}$M) algorithm by 
extending a single particle pair error measure which has been given by 
Hockney and Eastwood.
We also derive an easy-to-use analytic approximation to the error formula.
This allows a straightforward and precise determination of the optimal 
splitting parameter (as a function of system specifications and P$^{3}$M 
parameters) and hence knowledge of the force accuracy prior to the actual 
simulation.
The high quality of the estimate is demonstrated in several examples.
}
}
\end{centering}
\end{abstract}



\narrowtext


\section*{Introduction}

The combination of periodic boundary conditions and long range
interactions is a frequent difficulty encountered in computer
simulations of physical systems, and the ingenious summation
technique connected with the name of Ewald\cite{Ewald} has 
become a standard instrument for tackling this problem.
However, it has long been realized that for rather extensive
simulations (involving, e.g., many particles) this approach is
still too time consuming and various alternative methods have
been invented. One particular class of these new algorithms
owes its speedup to an inspired replacement of the Fourier
transformation -- which lies at the heart of the Ewald 
technique -- by Fast Fourier (FFT) routines\cite{HE,DYP,EPBDLP}. 

In a recent publication\cite{DH} we presented a unified view 
of these FFT accelerated Ewald sums and carried out detailed 
accuracy measurements. 
However, since all these algorithms contain various free parameters, 
working at the maximally obtainable accuracy requires the user to 
tune them very carefully. This is straightforward if there exists 
a theo\-re\-ti\-cal estimate of the errors involved -- as is 
the case for the standard Ewald sum\cite{KP} as well as for the 
so-called particle mesh Ewald (PME) method\cite{P} -- but rather 
tedious otherwise.

In this article we present such an estimate for the root 
mean square error in the force of the so-called 
par\-ti\-cle-par\-ti\-cle--par\-ti\-cle-mesh (P$^{3}$M) 
algorithm by extending an error measure already derived by 
Hockney and Eastwood\cite{HE} and additionally provide an 
easy to use analytical approximation to the somewhat unwieldy 
expression comprising various sums.


\section*{Ewald sum and P$^{3}$M in a nutshell}

In this section we outline for reference purpose the most important 
formulas for the P$^{3}$M method without much explanation, derivation 
or motivation. For the details the reader is referred to the original 
P$^{3}$M literature\cite{HE} as well as our previous publication\cite{DH}.

\begin{figure}[t]
\vspace{3.1cm}
\end{figure}

Consider a system of $N$ particles with charges $q_{i}$ at positions 
$\VECr_{i}$ in an overall neutral cubic simulation box of length $L$.
Employing the Ewald sum, the force $\VECF_{i}$ on particle $i$, which 
results from all interactions with the other charges (including all 
periodic images), can be written in the following way:
\begin{equation}\label{EwaldKraftAnteile}
\VECF_{i} = \VECF_{i}^{(r)} + \VECF_{i}^{(k)} + \VECF_{i}^{(d)}
\end{equation}
The so-called real space, Fourier space and dipole contributions 
are respectively given by
\begin{eqnarray}
\VECF_{i}^{(r)} & = & q_{i} \sum_{j} q_{j} \sum_{\VECm\in\ZZ^{3}}^{\prime}
\bigg( \frac{2\alpha}{\sqrt{\pi}}\exp(-\alpha^{2}|\VECr_{ij}+\VECm L|^{2}) 
+ \nonumber \\
{~} & {~} & + \frac{\erfc(\alpha|\VECr_{ij}+\VECm L|)}{|\VECr_{ij}+\VECm L|}
\bigg) \frac{\VECr_{ij}+\VECm L}{|\VECr_{ij}+\VECm L|^{2}}
\label{Realraumanteil_Kraft} \\
\VECF_{i}^{(k)} & = & \frac{q_{i}}{L^{3}}\sum_{j}q_{j}\sum_{\VECk\ne 0}
\frac{4\pi\VECk}{k^{2}}\exp\left(-\frac{k^{2}}{4\alpha^{2}}\right)
\sin(\VECk\cdot\VECr_{ij}) \label{Impulsraumanteil_Kraft} \\
\VECF_{i}^{(d)} & = & -\frac{4\pi q_{i}}{(1+2\epsilon')L^{3}}
\sum_{j}q_{j}\VECr_{j} \label{Dipolanteil_Kraft}
\end{eqnarray}
The prime on the second sum in Eqn.\ (\ref{Realraumanteil_Kraft}) 
indicates that for $i=j$ the term $\VECm=0$ has to be omitted, $\erfc(r) 
:= 2\pi^{-1/2}\int_{r}^{\infty}\infd t \exp(-t^{2})$ is the complementary 
error function and of course $\VECr_{ij}=\VECr_{i}-\VECr_{j}$.
Furthermore, the inverse length $\alpha$ is the {\em splitting parameter} 
of the Ewald sum, which controls the relative importance of the contributions 
coming from real and reciprocal space, the $\VECk$-vectors are from the 
discrete set $\frac{2\pi}{L}\ZZ^{3}$ and $\epsilon'$ is the dielectric constant 
of the medium, which surrounds the cluster of simulation boxes as it tends 
(in a spherical way) towards an infinite system. 
In practice, the infinite sums in Eqns.\ (\ref{Realraumanteil_Kraft},
\ref{Impulsraumanteil_Kraft}) are truncated by only taking into account 
distances which are smaller than some real space cutoff $r\maxi$ and wave 
vectors with a modulus smaller than some reciprocal space cutoff $k\maxi$.

The P$^{3}$M method offers a fast way for an approximate computation
of the reciprocal space contribution (\ref{Impulsraumanteil_Kraft}). 
By mapping the system onto a mesh, the necessary Fourier transformations 
can be accomplished by Fast Fourier routines. At the same time the 
simple Coulomb Green function $4\pi/k^{2}$ is adjusted as to make the 
result of the mesh calculation most closely resemble the con\-ti\-nu\-um 
solution. 

The first step, i.e., generating the mesh based charge density 
$\rho\Mesh$ (defined at the mesh points $\VECr_{p}$), is carried
out with the help of a charge assignment function $W$:
\begin{equation}\label{chargeassignment3d_}
\rho\Mesh(\VECr_{p}) = \frac{1}{h^{3}} \sum_{i=1}^{N} 
q_{i} W(\VECr_{p}-\VECr_{i}) 
\end{equation} 
Here $h$ is the mesh spacing, and the number of mesh points 
$N\Mesh=L/h$ along each direction should preferably be a power of 
two, since in this case the FFT is most efficient.
The charge assignment function is classified according to its order 
$P$, i.e.\ between how many grid points -- per coordinate direction -- 
each charge is distributed. In the P$^{3}$M method introduced by 
Hockney and Eastwood\cite{HE} its Fourier transform is
\begin{equation}
\tilde{W}(\VECk) = 
h^{3} \,\left(
\frac{\sin(\frac{1}{2}k_{x}h)}{\frac{1}{2}k_{x}h} \,
\frac{\sin(\frac{1}{2}k_{y}h)}{\frac{1}{2}k_{y}h} \,
\frac{\sin(\frac{1}{2}k_{z}h)}{\frac{1}{2}k_{z}h}
\right)^{P}
\end{equation}

In a second step the mesh based electric field $\VECE(\VECr_{p})$ 
is calculated. Basically, the electric field is the derivative of 
the electrostatic potential, but there exist several alternatives
for implementing the differentiation on a lattice\cite{DH}. In this 
article we will restrict to the case of $i\VECk${\em-differentiation}, 
which works by multiplying the Fourier transformed potential with 
$i\VECk$. In this case $\VECE(\VECr_{p})$ can be written as
\begin{equation}\label{MeshEfeld}
\VECE(\VECr_{p}) \; = \;\; \stackrel{\longleftarrow}{\mbox{FFT}}
\left[ -i\VECk \;\times\; 
\stackrel{\longrightarrow}{\mbox{FFT}}\left[\rho\Mesh\right]
\;\times\;\hat{G}\opt 
\right](\VECr_{p}) 
\end{equation}
In words, $\VECE(\VECr_{p})$ is the {\em backward} finite Fourier
transform of the product of $-i\VECk$, the {\em forward} finite
Fourier transform of the mesh based charge density $\rho\Mesh$ 
and the so-called optimal influence function $\hat{G}\opt$, given by
\begin{equation}\label{optimalinfluencefunction}
\hat{G}\opt(\VECk) = 
\frac{\tilde{\VECD}(\VECk)\cdot\sum_{\VECm\in\ZZ^{3}}
\tilde{U}^{2}(\VECk+\frac{2\pi}{h}\VECm)\tilde{\VECR}
(\VECk+\frac{2\pi}{h}\VECm)}
{|\tilde{\VECD}(\VECk)|^{2}\left[\sum_{\VECm\in\ZZ^{3}}
\tilde{U}^{2}(\VECk+\frac{2\pi}{h}\VECm)\right]^{2}}
\end{equation}
\begin{eqnarray}
{\mbox{with}}\qquad\tilde{\VECR}(\VECk) & := & 
-i\VECk \frac{4\pi}{k^{2}} e^{-k^{2}/4\alpha^{2}} \label{referenceforce} \\
{\mbox{and}}\qquad\tilde{U}(\VECk) & := &
\tilde{W}(\VECk)/h^{3}\label{W->U}
\end{eqnarray}
Here $\tilde{\VECD}(\VECk)$ is the Fourier transform of the employed
differentiation operator, which is simply $i\VECk$ in our case.
Finally, one arrives at the force on particle $i$, i.e.\ the replacement 
of Eqn.\ (\ref{Impulsraumanteil_Kraft}):
\begin{equation}\label{ikKraft}
\VECF_{i} = q_{i} \sum_{\VECr_{p}\in\MM} \VECE(\VECr_{p}) 
W(\VECr_{i}-\VECr_{p})
\end{equation}
Hereby the sum extends over the complete mesh $\MM$.

Although the presented formulas (\ref{chargeassignment3d_}--\ref{ikKraft})
look somewhat complicated, it is rather easy to implement them step by 
step. Furthermore, due to the replacement of the Fourier transforms by FFT 
routines (see Eqn.\ (\ref{MeshEfeld})), the algorithm is not only fast but 
its CPU time shows a favorable scaling with particle number: If the real 
space cutoff $r\maxi$ is chosen small enough (so that the real space 
contribution (\ref{Realraumanteil_Kraft}) can be calculated in order $N$), 
the complete algorithm is essentially of order $N\log N$.


\section*{Scaling of the rms force error}

In this section we address the dependence of the root mean square 
error in the force on the number of charged particles and their 
valence. Since the assumptions and arguments involved are of a 
rather general nature, the result is not specific to a certain kind 
of Ewald method.

We define the rms error in the force to be
\begin{equation}\label{rmsforce}
\Delta F := \sqrt{\frac{1}{N}\sum_{i=1}^{N}
\left(\VECF_{i}-\VECF^{\mbox{\footnotesize exa}}_{i}\right)^{2}}
=: \sqrt{\frac{1}{N}\sum_{i=1}^{N}\left(\Delta\VECF_{i}\right)^{2}}
\end{equation}
where $\VECF_{i}$ is the force on particle $i$ calculated by the 
algorithm under investigation and $\VECF^{\mbox{\footnotesize exa}}_{i}$ 
is the {\em exact} force on that particle.
Note that this is by no means the only interesting measure of accuracy. 
However, it is the only one which is considered in this article.

We now assume that the error in the force on particle $i$ can be 
written as
\begin{equation}\label{DeltaFform}
\Delta\VECF_{i} = q_{i}\sum_{j\ne i}q_{j}\,\VECchi_{ij}
\end{equation}
The idea behind this ansatz is that -- just as it is true for $\VECF_{i}$ 
-- the {\em error} in $\VECF_{i}$ originates from the $N-1$ interactions 
of particle $i$ with the other charged particles, and each contribution 
should be proportional to the product of the two charges involved. The
vector $\VECchi_{ij}$ gives the direction and magnitude of this error for
two unit charges and depends on their separation and orientation as
well as on the specific algorithm used for calculating the electrostatic
forces. For this term we further assume
\begin{equation}\label{uncorrelated}
\left\langle\VECchi_{ij}\cdot\VECchi_{ik}\right\rangle = 
\delta_{jk}\left\langle \VECchi_{ij}^{2}\right\rangle =: \delta_{jk}\,\chi^{2}
\end{equation}
where averaging over the particle configurations is denoted by the angular 
brackets. The underlying assumption that contributions from different 
particles are uncorrelated is certainly not always true (think, e.g., of 
highly ordered or strongly inhomogeneous particle distributions), but it 
is sensible for {\em random} systems. 
Obviously, the term $\left\langle\VECchi_{ij}^{2}\right\rangle$ -- the mean
square force error for two unit charges -- can no longer depend on $i$ and 
$j$ and is thus written as $\chi^{2}$.
Using Eqns.\ (\ref{DeltaFform},\ref{uncorrelated}), it follows
\begin{equation}\label{rmszwi}
\left\langle (\Delta\VECF_{i})^{2} \right\rangle = 
q_{i}^{2} \sum_{j\ne i}\sum_{k\ne i} q_{j}q_{k}\langle \VECchi_{ij}\cdot\VECchi_{ik}
\rangle \approx q_{i}^{2}\,\chi^{2}\,\CALQ^{2}
\end{equation}
where the important quantity $\CALQ^{2}$ is defined as
\begin{equation}
\CALQ^{2} := \sum_{j=1}^{N} q_{j}^{2}
\end{equation}

Not all particles necessarily have the same charge. 
More specifically, let there be $P$ subsets $\NN_{p}$, defined by the 
condition that all $|\NN_{p}|$ particles from the subset $\NN_{p}$ 
have the same charge $c_{p}$. If $|\NN_{p}|\gg 1$, the law of 
large numbers and Eqn.\ (\ref{rmszwi}) gives
\begin{equation}
\frac{1}{|\NN_{p}|}\sum_{i\in\NN_{p}}(\Delta\VECF_{i})^{2} 
\approx \left\langle (\Delta\VECF_{i})^{2} \right\rangle_{i\in\NN_{p}}
\approx c_{p}^{2}\,\chi^{2}\,\CALQ^{2}
\end{equation}
i.e., the {\em arithmetic mean} of the $(\Delta\VECF_{i})^{2}$ for 
all particles $i\in\NN_{p}$ can be approximated by the {\em ensemble 
average} for {\em one} particle from $\NN_{p}$. 
In the case where {\em all} $|\NN_{p}|$ are large, it follows
\begin{eqnarray}
\frac{1}{N}\sum_{i=1}^{N} \left(\Delta\VECF_{i}\right)^{2} & = &
\sum_{p=1}^{P}\frac{|\NN_{p}|}{N}\left(\frac{1}{|\NN_{p}|}
\sum_{i\in\NN_{p}} (\Delta\VECF_{i})^{2} \right) \nonumber \\
~ & \approx & \chi^{2}\,\frac{\CALQ^{2}}{N}\sum_{p=1}^{P} |\NN_{p}| c_{p}^{2}
= \chi^{2}\,\frac{\CALQ^{4}}{N}
\end{eqnarray}
Inserting this into Eqn.\ (\ref{rmsforce}) gives the final relation
\begin{equation}\label{rmsscale}
\Delta F \approx \chi\,\frac{\CALQ^{2}}{\sqrt{N}}
\end{equation}
Thus, the scaling of the rms error in the force with particle number
and valence is given by the factor $\CALQ^{2}N^{-1/2}$, whereas the
prefactor $\chi$ -- which cannot be obtained by such simple arguments --
contains the details of the method. Indeed, the estimates for the
real and reciprocal space error of the standard Ewald sum\cite{KP} 
as well as the estimate for the reciprocal space error of the PME 
method\cite{P} are exactly of the form (\ref{rmsscale}). 
Note that any information on the valence distribution enters only 
through the value of $\CALQ^{2}$.


\section*{The error measure of Hockney and Eastwood}

The most interesting ingredient of the P$^{3}$M method is the optimal
influence function from Eqn.\ (\ref{optimalinfluencefunction}). It is
constructed such that the result of the mesh calculation is as close as
possible to the solution of the original continuum problem. Of course,
this is only realizable in a quantitative way, if the notion of ``as 
close as possible'' is stated more precisely.
Hockney and Eastwood define the following measure of the error involved
in a P$^{3}$M calculation:
\begin{equation}\label{P3MQ}
Q := \frac{1}{h^{3}}\int_{h^{3}}\infd^{3}r_{1}\int_{L^{3}}\infd^{3}r
\big[\VECF(\VECr;\VECr_{1})-\VECR(\VECr)\big]^{2}
\end{equation}
$\VECF(\VECr;\VECr_{1})$ is the Fourier space contribution of the force 
between two unit charges at positions $\VECr_{1}$ and $\VECr_{1}+\VECr$ as 
calculated by the P$^{3}$M method (note that due to broken rotational and 
translational symmetry this does in fact depend on the coordinates of {\em 
both} particles), and $\VECR(\VECr)$ is the corresponding exact reference 
force (whose Fourier transform is just Eqn.\ (\ref{referenceforce})).
The inner integral over $\VECr$ scans all particle separations, whereas the 
outer integral over $\VECr_{1}$ averages over all possible locations of the 
first particle within a mesh cell.
Obviously, up to a factor $L^{-3}$ this expression is just the mean
square error in the force for two unit charges, in other words, the
quantity $\chi^{2}$ from Eqn.\ (\ref{uncorrelated}). This provides a 
link between the rms error of an $N$ particle system and the error $Q$ 
from Hockney and Eastwood: Using Eqn.\ (\ref{rmsscale}) one obtains
\begin{equation}\label{P3Mrms}
\Delta F \approx \CALQ^{2}\sqrt{\frac{Q}{N L^{3}}}
\end{equation}

Technically spoken, $Q$ is a {\em functional} of the influence function, 
and by setting the functional derivative of $Q$ with respect to $\hat{G}$ 
to zero Hockney and Eastwood were able to derive the optimal influence
function from Eqn.\ (\ref{optimalinfluencefunction}).
However, it is most important to realize that they also provide a closed 
expression for the corresponding ``optimal error'' $Q\opt=Q[\hat{G}\opt]$:
\begin{eqnarray}\label{P3MQopt}
Q\opt & = & \frac{1}{L^{3}}\,\sum_{\VECk\in\hat{\MM}}
\Bigg\{
\sum_{\VECm\in\ZZ^{3}}
|\tilde{\VECR}(\VECk+\frac{2\pi}{h}\VECm)|^{2} - \nonumber \\
~ & - &  \frac{\Big|\tilde{\VECD}(\VECk)\cdot\sum_{\VECm\in\ZZ^{3}}
\tilde{U}^{2}(\VECk+\frac{2\pi}{h}\VECm)
\tilde{\VECR}^{\ast} (\VECk+\frac{2\pi}{h}\VECm)\Big|^{2}}
{|\tilde{\VECD}(\VECk)|^{2}\left[\sum_{\VECm\in\ZZ^{3}}
\tilde{U}^{2}(\VECk+\frac{2\pi}{h}\VECm)   \right]^{2}}
\Bigg\}
\end{eqnarray}
The outer sum extends over all $\VECk$-vectors of the Fourier 
transformed mesh $\hat{\MM}$, and the star denotes complex 
conjugation. 
Once again, in the special case of $i\VECk$-differentiation 
one has $\tilde{\VECD}(\VECk)=i\VECk$.

Admittedly, Eqn.\ (\ref{P3MQopt}) looks rather complicated. Still, in 
combination with Eqn.\ (\ref{P3Mrms}) it gives the rms force error of 
the P$^{3}$M method (or -- more precisely -- of its Fourier space 
contribution)! (After all, the computation of $Q\opt$ and that of 
$\hat{G}\opt$ are quite similar.)
We would like to emphasize that the formula (\ref{P3MQopt}) for the 
optimal $Q$-value (just like the one for the optimal influence function
(\ref{optimalinfluencefunction})) is of a very general nature: It does
also work for different charge assignment functions, reference forces
or any differentiation scheme which can be expressed by an operator 
$\tilde{\VECD}(\VECk)$, in particular for all the finite difference 
schemes presented in our previous publication\cite{DH}.

The corresponding rms error in the force from the real space contribution
(\ref{Realraumanteil_Kraft}) has been derived by Kolafa and Perram\cite{KP}
and we want to provide it here for reference purpose:
\begin{equation}\label{realspaceerror}
\Delta F^{(r)} \approx \frac{2\CALQ^{2}}{\sqrt{Nr_{\mbox{\scriptsize max}} L^{3}}}\,
\exp(-\alpha^{2}r_{\mbox{\scriptsize max}}^{2})
\end{equation}
With these two estimates at hand it is easy to determine the optimal value 
of the splitting parameter $\alpha$ via a stand-alone program, which takes 
the relevant system parameters ($N$, $\CALQ^{2}$, $L$) and specifications 
of the algorithm ($r\maxi$, $N\Mesh$, $P$) as its input. 
If real and reciprocal space contribution to the error, $\Delta F^{(r)}$ 
and $\Delta F^{(k)}$ respectively, are assumed to be statistically 
independent, the total error is given by 
\begin{equation}
\Delta F = \sqrt{\left(\Delta F^{(r)}\right)^{2}+\left(\Delta F^{(k)}\right)^{2}}
\end{equation}
This quantity has to be minimized with respect to $\alpha$. However,
in most cases it is accurate enough to use the following approximation:
Determine the value of $\alpha$ at which the real and reciprocal space
contribution to the rms force error are equal. 


\section*{Analytic approximation}

Although the closed expression for the error from the last section
is not really {\em complicated}, it is somewhat {\em unwieldy}.
A possible calculation of the optimal value of $\alpha$ using e.g.\ 
a bisection method needs several computations of $Q\opt$ and for
each it is necessary to compute the inner aliasing sums and the outer 
sum over the $\VECk$-vectors. Especially for large Fourier meshes this 
can be rather time consuming. 
Therefore we now derive an analytic approximation to this error 
estimate, which is essentially an {\em expansion for small} $h\alpha$. 
We will restrict to the case of a cubic system and the same number 
$N\Mesh$ of mesh points along each direction. Also, we will only treat
the case of $i\VECk$-differentiation (see Eqn.\ (\ref{MeshEfeld})), since
we found\cite{DH} that this is the most accurate method. However, our
line of reasoning can be extended to more general cases. 

We start our treatment of Eqn.\ (\ref{P3MQopt}) by observing that two of 
the three sums over $\ZZ^{3}$ contain $\tilde{\VECR}$ with its exponential 
factor $\exp(-k^{2}/4\alpha^{2})$. Since near the boundary of $\hat{\MM}$ 
its value is roughly given by $\exp(-(\pi/h\alpha)^{2})$, $\tilde{\VECR}$ 
is strongly damped outside $\hat{\MM}$ if $h\alpha$ is small. Thus, it is 
a good approximation to retain only the term with $\VECm=0$ in these two 
sums. Inserting $\tilde{\VECD}(\VECk)=i\VECk$, the Fourier transform of the 
charge assignment function from Eqn.\ (\ref{chargeassignment3d_}) and 
using the fact that $\sin^{2}(x+n\pi)=\sin^{2}(x)$ for integral $n$, one 
obtains
\begin{eqnarray}
Q\opt & \approx &
\frac{(4\pi)^{2}}{L^{3}}\sum_{\VECk\in\hat{\MM}}
\frac{e^{-k^{2}/2\alpha^{2}}}{k^{2}} \times \nonumber \\
~ & \times & \left\{1- f^{(P)}(\frac{k_{x}h}{2}) f^{(P)}(\frac{k_{y}h}{2}) 
f^{(P)}(\frac{k_{z}h}{2}) \right\} \label{Qapprox1}
\end{eqnarray}
with the function $f^{(P)}$ defined as
\begin{equation}\label{fdef}
f^{(P)}(x) := \frac{x^{-4P}}
{\left(\sum_{m\in\ZZ}\;(x+m\pi)^{-2P}\right)^{2}}
\end{equation}
In a second step the aliasing sum in the denominator of 
Eqn.\ (\ref{fdef}) is evaluated analytically by exploiting 
the following partial fraction expansion\cite{AS}:
\begin{equation}\label{Kotangensidentitaet}
\sin^{-2}(x) = \sum_{m\in\ZZ} \, (x+m\pi)^{-2}
\quad,\quad x\in\RR \backslash \pi\ZZ
\end{equation}
Differentiating this expression $2P-2$ times gives
\begin{equation}
\sum_{m\in\ZZ} \, (x+m\pi)^{-2P} = 
\frac{1}{(2P-1)!}\frac{\infd^{2P-2}}{\infd x^{2P-2}}\sin^{-2}(x)
\end{equation}
This equation leads to a closed expression for the function
$f^{(P)}$ from Eqn.\ (\ref{fdef}):
\begin{equation}\label{fclosed}
f^{(P)}(x) = \frac{\big[(2P-1)!\big]^{2}}{x^{4P}}\,
\left(\frac{\infd^{2P-2}}{\infd x^{2P-2}}\sin^{-2}(x)\right)^{-2}
\end{equation}
Unfortunately the sum over $\hat{\MM}$ is still too complicated
to perform, so some further approximations are necessary. 
We choose the following way:
$f^{(P)}$ is expanded in a Taylor series up to order $4P-2$.
Since {\em (i)} $f^{(P)}$ is an even function, {\em (ii)} 
$f^{(P)}(0)=1$ and {\em (iii)} the lowest nontrivial term is of 
order $x^{2P}$, this expansion can be written as
\begin{equation}\label{fexpansion}
f^{(P)}(x) \approx f_{T}^{(P)}(x) :=
1 - x^{2P} \, \sum_{m=0}^{P-1} \, c_{m}^{(P)} \, x^{2m}
\end{equation}
The coefficients $c_{m}^{(P)}$ are easily determined with the
help of any mathematical computer program capable of symbolic
algebra and are listed in Table \ref{ctable}.
The term in curly brackets from Eqn.\ (\ref{Qapprox1}) 
can now be approximated like this:
\[
1-f^{(P)}(\frac{k_{x}h}{2})
  f^{(P)}(\frac{k_{y}h}{2})
  f^{(P)}(\frac{k_{z}h}{2}) \approx 
\]
\begin{equation}\label{factorapprox}
\Big(1-f_{T}^{(P)}(\frac{k_{x}h}{2})\Big) +
\Big(1-f_{T}^{(P)}(\frac{k_{y}h}{2})\Big) +
\Big(1-f_{T}^{(P)}(\frac{k_{z}h}{2})\Big)
\end{equation}
The product of the three functions $f^{(P)}$ is computed by
multiplying their Taylor expansions term by term, but the results
are only retained up to the truncation order $4P-2$. Note that 
the first neglected cross term would be of order $4P$.

For symmetry reasons it is clear that all three terms in the last 
line of (\ref{factorapprox}) contribute in the same way to the 
value of the sum in Eqn.\ (\ref{Qapprox1}), therefore it suffices
to choose one of them -- e.g., the $z$-term -- and multiply the 
result by 3. Together with the definition of $f_{T}^{(P)}$ from 
Eqn.\ (\ref{fexpansion}) this leads to
\begin{equation}
Q\opt \approx
3\,\frac{(4\pi)^{2}}{L^{3}}\sum_{\VECk\in\hat{\MM}}
\frac{e^{-k^{2}/2\alpha^{2}}}{k^{2}}\sum_{m=0}^{P-1}c_{m}^{(P)}
\left(\frac{k_{z}h}{2}\right)^{2(P+m)}
\end{equation}
Finally, the sum is replaced by an integral via
\begin{equation}
\left(\frac{2\pi}{L}\right)^{3}\sum_{\VECk} \longrightarrow \int\infd^{3}k
\end{equation}
If one extends the range of integration to $\RR^{3}$ and changes 
to spherical polar coordinates, the remaining angular and radial 
integrals can be performed with the help of
\begin{equation}
\int_{0}^{\pi}\infd\vartheta\,\sin\vartheta\,\cos^{2n}\vartheta = 
\frac{2}{2n+1} \quad,\quad n\in\NN
\end{equation}
and
\begin{equation}
\int_{0}^{\infty}\infd x \, x^{2n}e^{-x^{2}}=
\frac{\sqrt{\pi}\,(2n-1)!!}{2^{n+1}}\quad,\quad n\in\NN
\end{equation}
where $(2n-1)!!=1\cdot3\cdot5\cdots(2n-1)$. 
Collecting all parts together gives
\begin{equation}
Q\opt \approx
\sqrt{2\pi}\,\alpha\,\left(h\alpha\right)^{2P}\sum_{m=0}^{P-1}a_{m}^{(P)}\left(h\alpha\right)^{2m}
\end{equation}
with the abbreviation
\begin{equation}\label{aexpansion}
a_{m}^{(P)} := 12 \, \frac{\big(2(P+m)-1\big)!!}{2(P+m)+1}\,2^{-2(P+m)}\,c_{m}^{(P)}
\end{equation}

Combining this with Eqn.\ (\ref{P3Mrms}) results in the following
analytical approximation to the Fourier space contribution to
the rms force error of the P$^{3}$M algorithm:
\begin{equation}\label{mainresult}
\Delta F \approx
\frac{\CALQ^{2}}{L^{2}}\left(h\alpha\right)^{P}
\sqrt{\frac{\alpha L}{N}\sqrt{2\pi}\sum_{m=0}^{P-1}a_{m}^{(P)}\left(h\alpha\right)^{2m}}
\end{equation}
The exact expansion coefficients $a_{m}^{(P)}$ (which are rational numbers) 
are listed in Table \ref{atable}.

Let us repeat that Eqn.\ (\ref{mainresult}) was derived under the explicit
assumption that $h\alpha$ is small. Both, the restriction to the term 
$\VECm=0$ for two sums in Eqn.\ (\ref{P3MQopt}) as well as the expansion of 
the function $f^{(P)}$ from Eqns.\ (\ref{fdef},\ref{fclosed}) in powers of 
$h\alpha$ can become questionable if $h\alpha$ becomes large. 
However, in this case it is still safe to go back to the original error 
estimate, i.e.\ the combination of Eqns.\ (\ref{P3Mrms}) and (\ref{P3MQopt}). 


\section*{Numerical test}

\begin{figure}[t!]
\hspace{-1.1cm}\parbox{8.6cm}{\include{f1}}
\caption{The rms error $\Delta F$ (solid lines) for the 
system of 100 randomly distributed charges is calculated for 
the $i\VECk$-differentiated P$^{3}$M method with $N\Mesh=32$ 
mesh points and real space cutoff $r\maxi=4$. 
From top to bottom the order of the charge assignment function 
is increased from 1 to 7.
The dotted lines are the corresponding full estimates (using
Eqns.\ (\ref{P3Mrms},\ref{P3MQopt})) for the Fourier space 
contribution to $\Delta F$.}\label{rms_P_exa}
\vspace{1cm}
\hspace{-1.1cm}\parbox{8.6cm}{\include{f2}}
\caption{Same plot as in Fig.\ \ref{rms_P_exa}, but here the dotted 
lines are the analytical estimates from Eqn.\ (\ref{mainresult}).
Note that for large $P$ the error $\Delta F$ is overestimated
at large values of $\alpha$.}\label{rms_P_ana}
\end{figure}

In this section we demonstrate the accuracy of the P$^{3}$M error 
estimates by comparing their predictions with the exact rms force 
error $\Delta F$ from Eqn.\ (\ref{rmsforce}) -- calculated for a 
specific random system. Hereby the exact forces $\VECF_{i}$ needed
for computing $\Delta F$ are obtained by a well converged standard
Ewald sum, and for the test system we choose the one described in 
Appendix D of our previous publication\cite{DH}: 100 particles randomly 
distributed within a cubic box of length $L=10$, half of them carry a 
positive, the other half a negative unit charge. Our unit conventions 
are as follows\cite{DH}: lengths are measured in $\CALL$ and charges in 
$\CALC$. Hence the unit of force is $\CALC^{2}/\CALL^{2}$.
We will refer to the estimate which emerges from combining Eqns.\ 
(\ref{P3Mrms}) and (\ref{P3MQopt}) as the {\em full estimate}
and to Eqn.\ (\ref{mainresult}) as the {\em analytical approximation}.

\begin{figure}[t!]
\hspace{-1.1cm}\parbox{8.6cm}{\include{f3}}
\caption{$\Delta F$ (solid lines) for the same system as in the
previous figures is calculated for the $i\VECk$-differentiated 
P$^{3}$M method with charge assignment order $P=3$ and real space 
cutoff $r\maxi=4$. From top to bottom the number of mesh points 
varies like 4, 8, 16, 32, 64, 128. The dotted lines are the 
corresponding full error estimates.}\label{rms_M_exa}
\vspace{1cm}
\hspace{-1.1cm}\parbox{8.6cm}{\include{f4}}
\caption{Same plot as in Fig.\ \ref{rms_M_exa}, but here the dotted 
lines are the analytical estimates from Eqn.\ (\ref{mainresult}).
At small $N\Mesh$, which corresponds to large values of $h=L/N\Mesh$, 
the error formula overestimates $\Delta F$ at large values 
of $\alpha$.}\label{rms_M_ana}
\end{figure}

In a first example we fix the number of mesh points to $N\Mesh=32$ and 
the real space cutoff to $r\maxi=4$. The charge assignment order varies 
from $P=1$ through $P=7$. In Fig.\ \ref{rms_P_exa} the resulting curves 
for the rms force error $\Delta F$ are plotted together with the full 
error estimate and in Fig.\ \ref{rms_P_ana} the same is done for the 
analytical approximation.
It can be seen very clearly that the full estimate accurately predicts 
the Fourier space contribution to $\Delta F$ for all values of $\alpha$
and $P$. Since the real space contribution is also known\cite{KP} (see 
also Eqn.\ (\ref{realspaceerror})), this permits an easy determination 
of the optimal value of the splitting parameter $\alpha$.
The analytical approximation is almost as accurate as the full formula,
however, for large $P$ it diminishes in accuracy if $\alpha$ gets large. 
This is due to the fact that Eqn.\ (\ref{mainresult}) was derived under 
the assumption that $h\alpha$ is small. Note that the expansion coefficients
$a_{m}^{(P)}$ needed in Eqn.\ (\ref{mainresult}) strongly increase with
increasing $m$ if $P$ gets larger (see Table \ref{atable}). 
Still, both estimates are useful for determining the optimal operation point, 
and Eqn.\ (\ref{mainresult}) can be calculated much faster than the sums from 
Eqn.\ (\ref{P3MQopt}).

\begin{figure}[t!]
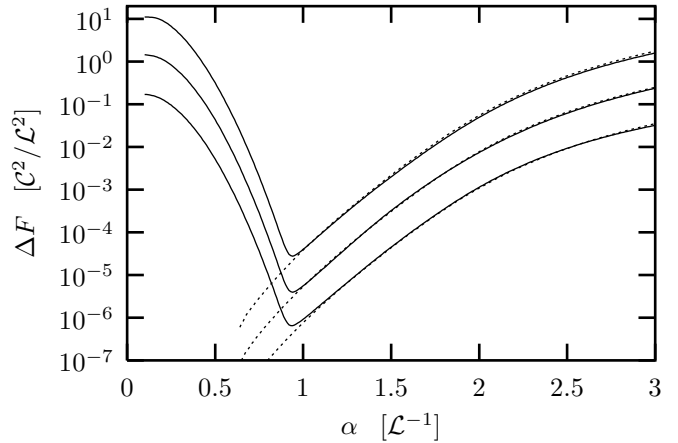

\hspace{-1.1cm}\parbox{8.6cm}{\include{f5}}
\caption{Test of the $\CALQ^{2}N^{-1/2}$ scaling of $\Delta F$. 
The three solid lines show the rms force error $\Delta F$ for 
systems, which differ in their $(\CALQ^{2},N)$ values. From top 
to bottom they are characterized by $(10000,400)$, $(1000,200)$ 
and $(100,100)$ (the last system is the same as the one in 
Figs.\ \ref{rms_P_exa}--\ref{rms_M_ana}).
The dotted curves are the corresponding full error 
estimates.}\label{rms_Q_exa}
\end{figure}

Next we study at fixed charge assignment order $P=3$ different mesh
sizes $h=L/N\Mesh$ by investigating $N\Mesh\in\{4,8,16,32,64,128\}$. 
Figs.\ \ref{rms_M_exa} and \ref{rms_M_ana} show $\Delta F$ in comparison 
with the full and the approximated error estimate respectively.
Again, it can be seen that the former very accurately gives the 
Fourier space contribution to $\Delta F$. As expected, the analytical 
approximation has problems at small $N\Mesh$ (since this results in large 
$h$), but nevertheless it is very useful otherwise. Essentially, one
has to check the value of $h\alpha$: If this is of the order unity or even 
larger, care is called for. Note that in Fig.\ \ref{rms_M_ana} the 
value of $h\alpha$ is approximately 1 at the points where the analytical
approximation starts to deviate from the true curve.

In a next step we want to demonstrate that the scaling of the rms force 
error with particle number and valence distribution is in fact correctly
given by $\Delta F \propto \CALQ^{2}N^{-1/2}$.
To this end we investigate three systems which differ {\em only} in the 
values of $\CALQ^{2}$ and $N$. The first system is the same as the one 
investigated so far in Figs.\ \ref{rms_P_exa}--\ref{rms_M_ana}. A second 
system contains 200 particles, namely, 50 monovalent and 50 trivalent 
pairs. Finally, a third system contains 400 particles: 50 pairs with 
charge $\pm 1$, 100 pairs with charge $\pm 5$ and 50 pairs with charge 
$\pm 7$. 
Hence, their $(\CALQ^{2},N)$ values are respectively given by $(100,100)$, 
$(1000,200)$ and $(10000,400)$, and the ratio of their scaling prefactors 
is thus $1:\sqrt{50}:50$.
In Fig.\ \ref{rms_Q_exa} this is clearly visible in the constant shift 
of the three curves with respect to one another (note that the vertical 
scale is logarithmic). Also shown is the full error estimate, which 
again predicts the Fourier space contribution to $\Delta F$ in all three 
cases very precisely.


\section*{Application to an inhomogeneous polyelectrolyte system}

\begin{figure}[t!]
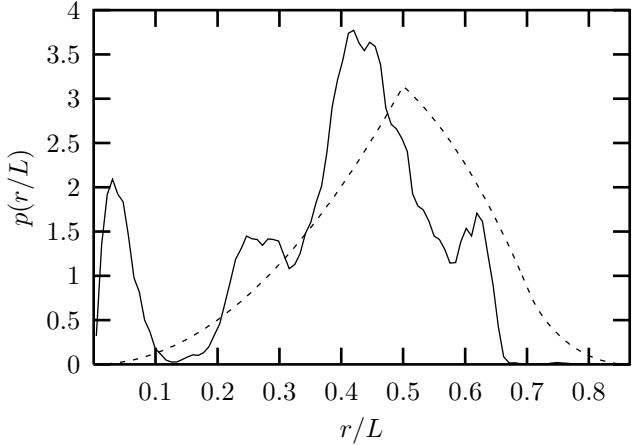

\hspace{-0.1cm}\parbox{8.6cm}{\include{f6}}
\caption{Measured relative frequency $p(r/L)$ for the scaled minimum 
image separation $r/L$ between two charges of the polyelectrolyte system 
described in the text (solid line). For comparison, the probability density 
of $r/L$ for a random system is also shown (dashed line).}\label{rms_inhomo_minimg}
\end{figure}

So far we have only used homogeneous random systems for testing
the error estimates. However, this does not necessarily reflect
the situation encountered in all computer experiments. In the 
present section we want to show that deviations from a random 
distribution, as they frequently occur in charged systems, in 
fact have a noticeable influence on the rms error.

We will use a typical system from our own research to demonstrate
this effect: a simple model of a polyelectrolyte solution\cite{SK,DHKM}.
106 Lennard--Jones particles were joined (by some bonding potential) 
to build up a polymer chain. Every third ``monomer'' was monovalently 
charged and 8 such chains together with 96 trivalent and oppositely
charged counterions, which make the complete system electrically 
neutral, were put in a cubic simulation box of length $L\approx 179$. 
The system was brought into the canonical state by means of a Molecular 
Dynamics simulation and a Langevin thermostat.

Under certain circumstances (e.g., at sufficiently low temperature and 
the appropriate density range) such polyelectrolyte chains collapse, and 
this happened to the described system. The chain sizes shrunk to much 
smaller values than for comparable neutral polymers and 90\% of the 
counterions were condensed within a distance of only two Lennard--Jones 
radii from the nearest chain. 

\begin{figure}[t!]
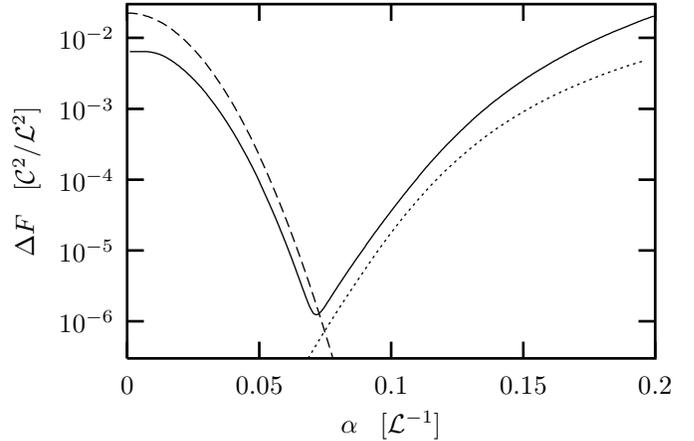

\hspace{-1.1cm}\parbox{8.6cm}{\include{f7}}
\caption{rms force error $\Delta F$ for the polyelectrolyte system 
described in the text (solid line). The dotted curve is the full 
P$^{3}$M estimate for the reciprocal space contribution, the dashed 
curve is the estimate for the real space contribution. Note that due 
to the strong inhomogeneities in the charge distribution (see Fig.\ 
\ref{rms_inhomo_minimg}) both estimates show systematic deviations 
from the true error curve.}\label{rms_inhomo_exa}
\end{figure}

The various phenomena leading to this transition, the influence of the 
system parameters or the dynamics are only a few of the interesting 
physical questions. However, the only thing which concerns us here is 
the fact that after the transition the system has developed local 
inhomogeneities. To demonstrate this we have plotted in Fig.\ 
\ref{rms_inhomo_minimg} the measured relative frequency $p(r/L)$ of the 
scaled {\em minimum image distance} $r/L \in [0;\sqrt{3}/2]$ between two 
charges (we did not distinguish between different valences) and compared 
this to the probability density of $r/L$ for a random system. 
The differences are indeed very pronounced. Apart from the more complicated
structure of the measured curve, note in particular that small separations 
are more frequent at the expense of larger ones.

For this system we calculated the rms force error $\Delta F$ and the 
corresponding estimates for the real and reciprocal space contribution 
(Eqns.\ (\ref{realspaceerror};\ref{P3Mrms},\ref{P3MQopt})). Since the
simulation box comprises 288 monovalent and 96 trivalent charges, we 
have $N=384$ and $\CALQ=1152$. 
The result is shown in Fig.\ \ref{rms_inhomo_exa}. No longer do the 
estimates correctly predict the two branches of $\Delta F$. Rather, 
at small values of $\alpha$ the algorithm gives better results than 
expected from Eqn.\ (\ref{realspaceerror}), while at large values of 
$\alpha$ the estimated $\Delta F$ is smaller than the actual one. 
However, this trend can be explained qualitatively in the following way:  
At small values of $\alpha$ the force (and also its error) is dominated 
by the real space contribution from Eqn.\ (\ref{Realraumanteil_Kraft}).
This error originates from neglected contributions beyond the real
space cutoff $r\maxi$. In the case of Fig. \ref{rms_inhomo_exa} we used 
$r\maxi/L\approx0.243$ and from Fig.\ \ref{rms_inhomo_minimg} it can
be seen that there are more particles within this cutoff (and thus less
beyond) than in the case of a random particle distribution, which should
lead to an enhanced real space accuracy. On the other hand, we 
demonstrated previously (see Fig. 7 of our recent publication\cite{DH})
that the rms error of the P$^{3}$M method strongly increases with decreasing 
minimum image distance. The general shift towards smaller $r$, which
can be observed in the polyelectrolyte system, should thus lead to an
enlarged reciprocal space error. Observable effects will occur at large 
values of $\alpha$, where this contribution to $\Delta F$ dominates.

Although the systematic deviations of the error estimates from the true 
curve are easily detectable, they are less dramatic than one could have 
guessed from a brief look at Fig.\ \ref{rms_inhomo_minimg}. 
The optimal splitting parameter from Fig.\ \ref{rms_inhomo_exa} is given 
by $\alpha\opt\approx 0.0715$ with a corresponding $\Delta F\opt\approx 
1.2 \times 10^{-6}$, while the intersection point of real and reciprocal 
space estimate occurs at $\alpha\approx 0.0740$, which predicts an error 
of $\Delta F \approx 9.3 \times 10^{-7}$. If the estimated value of 
$\alpha$ had been used, this would result in an error of $\Delta F\approx 
1.5 \times 10^{-6}$, which is roughly 25\% larger than at the optimal 
value of $\alpha$. If such a safety margin is considered already at the 
beginning, the {\em a priori} determination of the ``optimal'' value of 
$\alpha$ by means of Eqns.\ (\ref{P3Mrms},\ref{P3MQopt},\ref{realspaceerror}) 
together with an {\em a posteriori} validity check is still a good approach.

In any case, if one knows or at least has reasons to suspect that the 
investigated system is susceptible to the development of inhomogeneities, 
one should always be aware of a potential failure of the presented error 
formulas. In case of doubt, some simple numerical tests -- like, e.g., 
the ones which we have performed here -- are neither out of place nor costly.


\section*{Conclusions}

We have presented an accurate estimate of the root mean square error
in the force involved in a P$^{3}$M calculation and additionally 
derived an easy-to-use analytic approximation for the special case of
$i\VECk$-differentiation. Together with the existing estimate 
for the real space contribution to the rms force error this permits a 
determination of the optimal tuning parameter $\alpha$ and provides 
information on the accuracy which is to be expected.
It thus guarantees that one does not waste accuracy which could otherwise
be achieved at the same computational effort. Stated the other way around:
It prevents one from spending more computational effort for a desired
algorithmic accuracy than actually necessary.

In our previous publication\cite{DH} we showed that the
$i\VECk$-differentiated P$^{3}$M method is the most accurate algorithm for
the FFT accelerated Ewald sum. In combination with the error estimates this
gives a ``package'' for calculating electrostatic interactions in periodic 
boundary conditions which is easy to use, very precise, produces known and 
controllable errors and can thus be optimally tuned in advance.
If one wishes to rely on the discrete differentiation operators\cite{DH},
the full error formula (\ref{P3Mrms},\ref{P3MQopt}) will still work.
It is only in the case of the analytic differentiation scheme\cite{EPBDLP}
that none of the present error estimates is applicable.

As a last point, we stressed several times that the validity of the error 
estimates is subject to some additional requirements, concerning e.g.\ the 
homogeneity of the system. Nevertheless, it should be obvious that in any 
case a consultation of the error formulas -- perhaps only for a first 
starting point -- is superior to guessing the parameters or the use of 
$\alpha$-values which were historically handed down. 
The difficulties of the analytical approximation (\ref{mainresult}) at 
large values of $h\alpha$ are not a serious problem anyway, since in case 
of doubt one can always go back to the full estimate.


\section*{Acknowledgments}

Both authors are grateful to K.\ Kremer for encouragement and helpful 
comments. C.\ H.\ further thanks the DFG for financial support.
A large amount of computer time was generously provided by the HLRZ 
J{\"u}lich under grant hkf06.



\onecolumn

\begin{table}
\begin{tabular}{cccccccc}
$P$ & $c_{0}^{P}$ & $c_{1}^{P}$ & $c_{2}^{P}$ & $c_{3}^{P}$ & $c_{4}^{P}$ &
$c_{5}^{P}$ & $c_{6}^{P}$ \\[0.5ex] \hline \\
1 & $\D\frac{2}{3}$ & & & & & & \\[3ex]
2 & $\D\frac{2}{45}$ & $\D\frac{8}{189}$ & & & & & \\[3ex]
3 & $\D\frac{4}{945}$ & $\D\frac{2}{225}$ & $\D\frac{8}{1\,485}$ & & & & \\[3ex]
4 & $\D\frac{2}{4\,725}$ & $\D\frac{16}{10\,395}$ & $\D\frac{5\,528}{3\,869\,775}$ &
$\D\frac{32}{42\,525}$ & & & \\[3ex]
5 & $\D\frac{4}{93\,555}$ & $\D\frac{2\,764}{11\,609\,325}$ & $\D\frac{8}{25\,515}$ &
$\D\frac{7\,234}{32\,531\,625}$ & $\D\frac{350\,936}{3\,206\,852\,775}$ & & \\[3ex]
6 & $\D\frac{2\,764}{638\,512\,875}$ & $\D\frac{16}{467\,775}$ & $\D\frac{7\,234}{119\,282\,625}$
& $\D\frac{1\,403\,744}{25\,196\,700\,375}$ & $\D\frac{1\,396\,888}{40\,521\,009\,375}$ &
$\D\frac{2\,485\,856}{152\,506\,344\,375}$ & \\[3ex]
7 & $\D\frac{8}{18\,243\,225}$ & $\D\frac{7\,234}{1\,550\,674\,125}$ &
$\D\frac{701\,872}{65\,511\,420\,975}$ & $\D\frac{2\,793\,776}{225\,759\,909\,375}$ &
$\D\frac{1\,242\,928}{132\,172\,165\,125}$ & $\D\frac{1\,890\,912\,728}{352\,985\,880\,121\,875}$ &
$\D\frac{21\,053\,792}{8\,533\,724\,574\,375}$ \\[3ex]
\end{tabular}
\caption{Expansion coefficients $c_{m}^{(P)}$ for the functions 
  $f^{(P)}$ from Eqns.\ (\ref{fdef},\ref{fclosed}) as needed in 
  Eqn.\ (\ref{fexpansion}).}\label{ctable}
\end{table}


\begin{table}
\begin{tabular}{cccccccc}
$P$ & $a_{0}^{P}$ & $a_{1}^{P}$ & $a_{2}^{P}$ & $a_{3}^{P}$ & $a_{4}^{P}$ &
$a_{5}^{P}$ & $a_{6}^{P}$ \\[0.5ex] \hline \\
1 & $\D\frac{2}{3}$ & & & & & & \\[3ex]
2 & $\D\frac{1}{50}$ & $\D\frac{5}{294}$ & & & & & \\[3ex]
3 & $\D\frac{1}{588}$ & $\D\frac{7}{1\,440}$ & $\D\frac{21}{3\,872}$ & & & & \\[3ex]
4 & $\D\frac{1}{4\,320}$ & $\D\frac{3}{1\,936}$ & $\D\frac{7\,601}{2\,271\,360}$ &
$\D\frac{143}{28\,800}$ & & & \\[3ex]
5 & $\D\frac{1}{23\,232}$ & $\D\frac{7\,601}{13\,628\,160}$ & $\D\frac{143}{69\,120}$ &
$\D\frac{517\,231}{106\,536\,960}$ & $\D\frac{106\,640\,677}{11\,737\,571\,328}$ & & \\[3ex]
6 & $\D\frac{691}{68\,140\,800}$ & $\D\frac{13}{57\,600}$ & $\D\frac{47\,021}{35\,512\,320}$ &
$\D\frac{9\,694\,607}{2\,095\,994\,880}$ & $\D\frac{733\,191\,589}{59\,609\,088\,000}$ &
$\D\frac{326\,190\,917}{11\,700\,633\,600}$ & \\[3ex]
7 & $\D\frac{1}{345\,600}$ & $\D\frac{3\,617}{35\,512\,320}$ & $\D\frac{745\,739}{838\,397\,952}$
& $\D\frac{56\,399\,353}{12\,773\,376\,000}$ & $\D\frac{25\,091\,609}{1\,560\,084\,480}$ &
$\D\frac{1\,755\,948\,832\,039}{36\,229\,939\,200\,000}$ & $\D\frac{4\,887\,769\,399}{37\,838\,389\,248}$ \\[3ex]
\end{tabular}
\caption{Expansion coefficients $a_{m}^{(P)}$ from Eqn.\ (\ref{aexpansion}).}\label{atable}
\end{table}


\end{document}